\documentclass[twocolumn,english,3p]{elsarticle}
\usepackage[T1]{fontenc}
\usepackage[latin9]{inputenc}
\usepackage{verbatim}
\usepackage{float}
\usepackage{units}
\usepackage{amsmath}
\usepackage{amssymb}
\usepackage{graphicx}
\usepackage{esint}

\makeatletter
\def\Rm{\mathit{Rm}}

\renewcommand{\vec}[1]{\mbox{\boldmath $#1$}}

\usepackage{balance}

\makeatother

\usepackage{babel}
\begin{document}

\title{Feasible homopolar dynamo with sliding liquid-metal contacts}

\author[JP]{J\={a}nis Priede}

\ead{j.priede@coventry.ac.uk}

\author[RZ]{Raúl Avalos-Zúñiga}

\ead{ravalosz@ipn.mx}

\address[JP]{Applied Mathematics Research Centre, Coventry University, Coventry,
CV1 5FB, United Kingdom}

\address[RZ]{CICATA-Qro, Instituto Politécnico Nacional, Cerro Blanco 141, Colinas
del Cimatario, Querétaro, Mexico}
\begin{abstract}
We present a feasible homopolar dynamo design consisting of a flat,
multi-arm spiral coil, which is placed above a fast-spinning metal
ring and connected to the latter by sliding liquid-metal electrical
contacts. Using a simple, analytically solvable axisymmetric model,
we determine the optimal design of such a setup. For small contact
resistance, the lowest magnetic Reynolds number, $\Rm\approx34.6,$
at which the dynamo can work, is attained at the optimal ratio of
the outer and inner radii of the rings $R_{i}/R_{o}\approx0.36$ and
the spiral pitch angle $54.7{}^{\circ}.$ In a setup of two copper
rings with the thickness of $\unit[3]{cm},$ $R_{i}=\unit[10]{cm}$
and $R_{o}=\unit[30]{cm},$ self-excitation of the magnetic field
is expected at a critical rotation frequency around $\unit[10]{Hz}.$\end{abstract}
\maketitle

\section{Introduction}

The homopolar dynamo is one of the simplest models of the self-excitation
of magnetic field by moving conductors which is often used to illustrate
the dynamo action that is thought to be behind the magnetic fields
of the Earth, the Sun and other cosmic bodies \citep{Moffatt78,ARAA96}.
In its simplest form originally considered by Bullard \citep{Bullard-1955},
the dynamo consists of a solid metal disc which rotates about its
axis, and a wire twisted around it and connected through sliding contacts
to the rim and the axis of the disc. At a sufficiently high rotation
rate, the voltage induced by the rotation of the disc in the magnetic
field generated by an initial current perturbation can exceed the
voltage drop due to the ohmic resistance. At this point, initial perturbation
starts to grow exponentially leading to the self-excitation of current
and its associated magnetic.

This simple model has a number of important extensions and modifications.
For example, the Rikitake model \citep{Rikitake58} consisting of
two coupled disc dynamos is known to generate an oscillating magnetic
field with complex dynamics \citep{Plunian-1997}. The latter study
is based on a modified model using a radially sectioned disc with
azimuthal current added at the rim. This modification eliminates the
unphysical growth of the magnetic field in the limit of perfectly
conducting disc \citep{Moffatt-1979}. Using the same model we recently
showed that dynamo can be excited by the parametric resonance mechanism
at substantially reduced rotation rate when the latter contains harmonic
oscillations in certain frequency bands \citep{PAP10}.

Despite its simplicity no successful implementation of the disc dynamo
is known so far. The problem appears to be the sliding electrical
contacts which are required to convey the current between the rim
and the axis of the rotating disc. Electrical resistance of the sliding
contacts, usually made of solid graphite brushes, is typically by
several orders of magnitude higher than that of the rest of the setup.
This results in unrealistically high rotation rates which are required
for dynamo to operate \citep{RadRhe02}. To overcome this problem
we propose to use liquid-metal sliding electrical contacts similar
to those employed in homopolar motors and generators \citep{Maribo-etal10}.
The aim of this letter is to develop a feasible design of the disc
dynamo which could achieve self-excitation at realistic rotation rates.

\section{Physical and mathematical models}

\begin{figure}
\begin{centering}
\includegraphics[width=0.9\columnwidth]{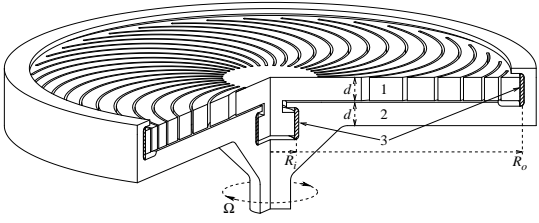} 
\par\end{centering}

\caption{\label{fig:sketch}Schematic view of the disc dynamo setup consisting
of a stationary coil (1) made of a copper disc sectioned by spiral
slits and a fast-spinning disc (2), which is electrically connected
to the former by sliding liquid-metal contacts (hatched) (3). }
\end{figure}

The principal setup of the proposed disc dynamo shown in Fig. \ref{fig:sketch}
consists of a stationary coil (1) made of a copper disc sectioned
by spiral slits and a fast-spinning disc (2) placed beneath it, which
is electrically connected to the former by sliding liquid electrical
contacts (3). The coil is supported by the holders which are not shown
in this basic view. The liquid metal is hold in vertical state by
the centrifugal force. Such a compact and symmetric design not only
minimizes the electrical resistance in the system but also makes it
amendable to simple analysis which is carried out in the following. 

To simplify the analysis both discs are subsequently assumed to be
thin coaxial rings of thickness $d,$ the outer radius $R_{o}\gg d,$
and the inner radius $R_{i}=\lambda R_{o},$ where $0<\lambda<1$
the ratio of the inner and outer radii. The rings are separated by
a small axial distance and connected to each other at their rims through
the sliding liquid-metal electrical contacts. The design of the stationary
top ring, which forms a compact coil consisting of spiral sections
is described in detail below. The bottom ring is mounted on an axle
which is driven by an electric motor with the angular velocity $\Omega.$
The electric current $I_{0}$ is induced by the rotation of the bottom
ring in the magnetic field generated by the same current returning
through the coil formed by the top ring. In the solid rotating ring,
the current is assumed to flow radially with the linear density $J_{r}=\frac{I_{0}}{2\pi r},$
which decreases due to the charge conservation inversely with the
cylindrical radius $r.$ Current returns through the top ring where
it is deflected by the spiral slits that produce an azimuthal component
proportional to the radial one: 
\begin{equation}
J_{\phi}=-J_{r}\beta=\frac{I_{0}\beta}{2\pi r},\label{eq:J-phi}
\end{equation}
where $\arctan\beta$ the pitch angle of the current lines relative
to the radial direction. The shape of slits following the current
lines is governed by $\frac{J_{\phi}}{J_{r}}=\frac{rd\phi}{dr}=-\beta$
and given by the logarithmic spirals 
\begin{equation}
\phi(r)=\phi_{0}-\beta\ln r,\label{eq:spiral}
\end{equation}
where $\phi$ is the azimuthal angle. The electric potential distribution
in the coil ring follows from Ohm's law 
\begin{equation}
\vec{J}=\frac{I_{0}}{2\pi r}\left(-\vec{e}_{r}+\beta\vec{e}_{\phi}\right)=-\sigma d\vec{\nabla}\varphi_{c},\label{eq:Jc}
\end{equation}
as 
\[
\varphi_{c}(r,\phi)=\frac{I_{0}}{2\pi\sigma d}\left(\ln r-\beta\phi\right).
\]
Thus, the potential difference along the current line between the
rims of the ring is 
\begin{equation}
\Delta\varphi_{c}=\left[\varphi_{c}(r,\phi(r))\right]_{R_{i}}^{R_{o}}=-\frac{I_{0}}{2\pi\sigma d}(1+\beta^{2})\ln\lambda.\label{eq:Phi-c}
\end{equation}

The potential difference across the bottom ring, which rotates as
a solid body with the azimuthal velocity $v_{\phi}=r\Omega,$ is defined
by the radial component of Ohm's law for a moving medium 
\[
J_{r}=\frac{I_{0}}{2\pi r}=\sigma d(-\partial_{r}\varphi_{d}+v_{\phi}B_{z}),
\]
where $B_{z}$ is the axial component of the magnetic field. Integrating
the expression above over the ring radius we obtain 
\begin{equation}
-\frac{I_{0}}{2\pi}\ln\lambda=\sigma d\left(-\Delta\varphi_{d}+\Omega\Phi_{d}\right),\label{eq:Phi-d}
\end{equation}
where $\Delta\varphi_{d}=\left[\varphi_{d}(r)\right]_{R_{i}}^{R_{o}}$
is the potential difference across the rotating ring and $\Phi_{d}=\int_{R_{i}}^{R_{o}}B_{z}r\, dr$
is the magnetic flux through it. Using the relation $B_{z}=r^{-1}\partial_{r}(rA_{\phi}),$
the latter can be expressed in terms of the azimuthal component of
the magnetic vector potential $A_{\phi}$ as 
\begin{equation}
\Phi_{d}=\left[rA_{\phi}\right]_{r=R_{i}}^{R_{o}}.\label{eq:flux-d}
\end{equation}
In the stationary state, which is assumed here, the potential difference
induced by the rotating ring in Eq. (\ref{eq:Phi-d}) is supposed
to balance that over the coil defined by Eq. (\ref{eq:Phi-c}) as
well as the potential drop over the liquid-metal contacts with the
effective resistance $\mathcal{R}:$ 
\begin{equation}
\Delta\varphi_{d}=\Delta\varphi_{c}+\mathcal{R}I_{0}.\label{eq:blnc}
\end{equation}
This equation implicitly defines the marginal rotation rate at which
a steady current can sustain itself.

\begin{figure*}
\begin{centering}
\includegraphics[width=0.5\textwidth]{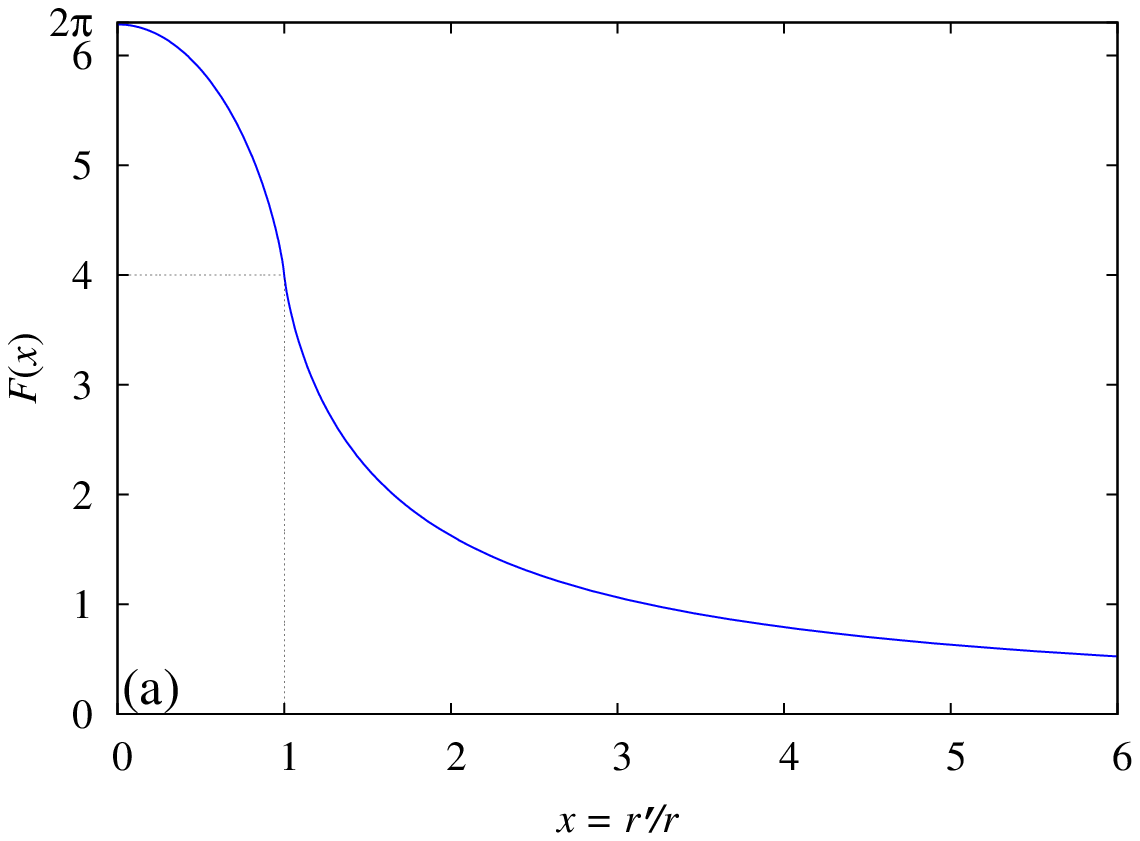}\includegraphics[width=0.5\textwidth]{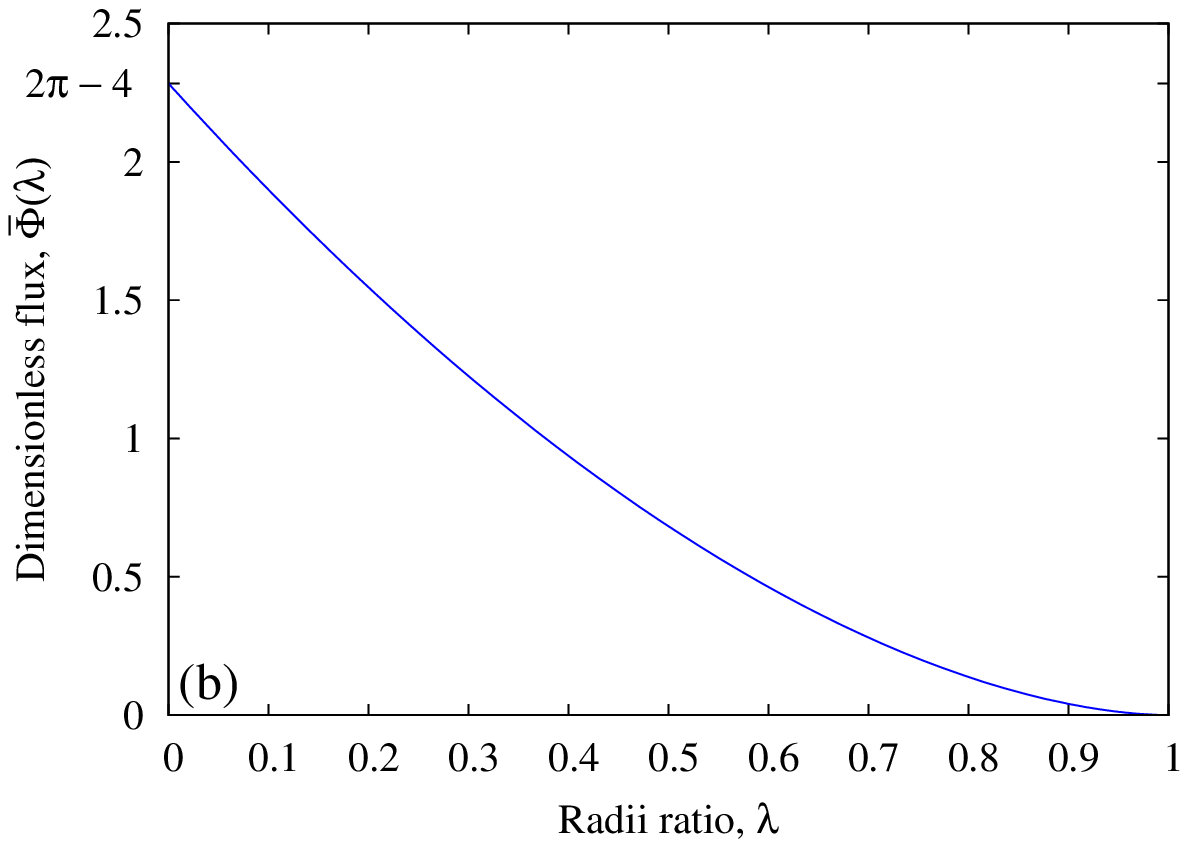} 
\par\end{centering}

\caption{\label{fig:flux-l}Function (\ref{eq:F(x)}) defining the vector potential
distribution in the plane of the coil (a) and the dimensionless magnetic
flux (\ref{eq:flux-l}) versus the radii ratio $\lambda$ (b). }
\end{figure*}

To complete the solution we need to evaluate the magnetic flux (\ref{eq:flux-d})
through the rotating disc. The azimuthal component of the vector potential
appearing in Eq. (\ref{eq:flux-d}) is generated by the respective
component of the electric current which is present only in the coil.
Thus, we have 
\[
A_{\phi}(r,z)=\frac{\mu_{0}}{4\pi}\intop_{0}^{2\pi}\intop_{R_{i}}^{R_{0}}\frac{J_{\phi}(r')\cos\phi\, r'\, dr'\, d\phi}{\sqrt{r'^{2}-2r'r\cos\phi+r^{2}+z^{2}}},
\]
where $z$ is the axial distance from the coil ring carrying the azimuthal
current $J_{\phi}$ defined by Eq. (\ref{eq:J-phi}). Note that the
poloidal currents with radial and axial components circulating through
the rings and liquid-metal contacts produce purely toroidal magnetic
field, which is parallel to the velocity of the rotating ring and,
thus, do not interact with the latter. In the plane of the ring $(z=0),$
the double integral above can be evaluated analytically as 
\[
A_{\phi}(r,0)=\frac{\mu_{0}\beta I_{0}}{8\pi^{2}}\left[F(R_{i}/r)-F(R_{o}/r)\right],
\]
where the function 
\begin{align}
F(x) & =(1-x)K(m_{+})+(1+x)E(m_{+}),\label{eq:F(x)}\\
 & +\mathrm{sgn}(1-x)\left[(1+x)K(m_{-})+(1-x)E(m_{-})\right]\nonumber 
\end{align}
which is produced by the computer algebra system Mathematica \citep{Mathematica}
in terms of the complete elliptic integrals of the first and second
kind, $K(m_{\pm})$ and $E(m_{\pm})$, of the \emph{parameter} $m_{\pm}=\frac{\pm4x}{(1\pm x)^{2}}$
\citep{AbSt72}, is plotted in Fig. \ref{fig:flux-l}(a). Taking into
account that $F(1)=4$, the magnetic flux (\ref{eq:flux-d}) can be
written as 
\begin{equation}
\Phi_{c}=\frac{\mu_{0}\beta I_{0}R_{o}}{8\pi^{2}}\bar{\Phi}(\lambda),\label{eq:flux-c}
\end{equation}
where 
\begin{equation}
\bar{\Phi}(\lambda)=F(\lambda)+\lambda F(\lambda^{-1})-4(1+\lambda)\label{eq:flux-l}
\end{equation}
is a dimensionless magnetic flux, which is plotted in Fig. \ref{fig:flux-l}(b)
versus the radii ratio $\lambda=R_{i}/R_{o}.$ 

In the following, we assume the axial separation between the rings
to be so small that the magnetic flux through the rotating ring is
effectively the same as that through the coil, i. e., $\Phi_{d}\approx\Phi_{c}.$
Substituting the relevant parameters into Eq. (\ref{eq:blnc}) we
eventually obtain 
\begin{equation}
\Rm=\mu_{0}\sigma dR_{o}\Omega=\frac{4\pi(\bar{\mathcal{R}}-(2+\kappa\beta^{2})\ln\lambda)}{\kappa\beta\bar{\Phi}(\lambda)},\label{eq:Rm}
\end{equation}
which is the marginal magnetic Reynolds number defining the dynamo
threshold depending on the spiral pitch angle $\arctan\beta,$ the
radii ratio $\lambda,$ and the dimensionless contact resistance $\bar{\mathcal{R}}=2\pi\sigma d\mathcal{R}.$
\begin{figure}[H]
\begin{centering}
\includegraphics[width=0.9\columnwidth]{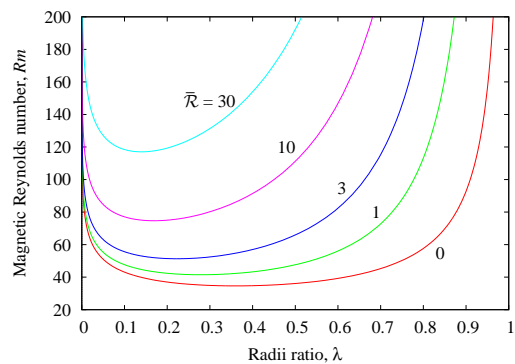} 
\par\end{centering}

\caption{\label{fig:Rm-lmb}Marginal $\Rm$ versus $\lambda$ for only one
ring sectioned $(\kappa=1)$ at various dimensionless contact resistances
$\bar{\mathcal{R}}$ and the optimal $\beta,$ which is plotted in
Fig. \ref{fig:crtpar}(b). }
\end{figure}
 
\begin{figure*}[!]
\begin{centering}
\includegraphics[width=0.5\textwidth]{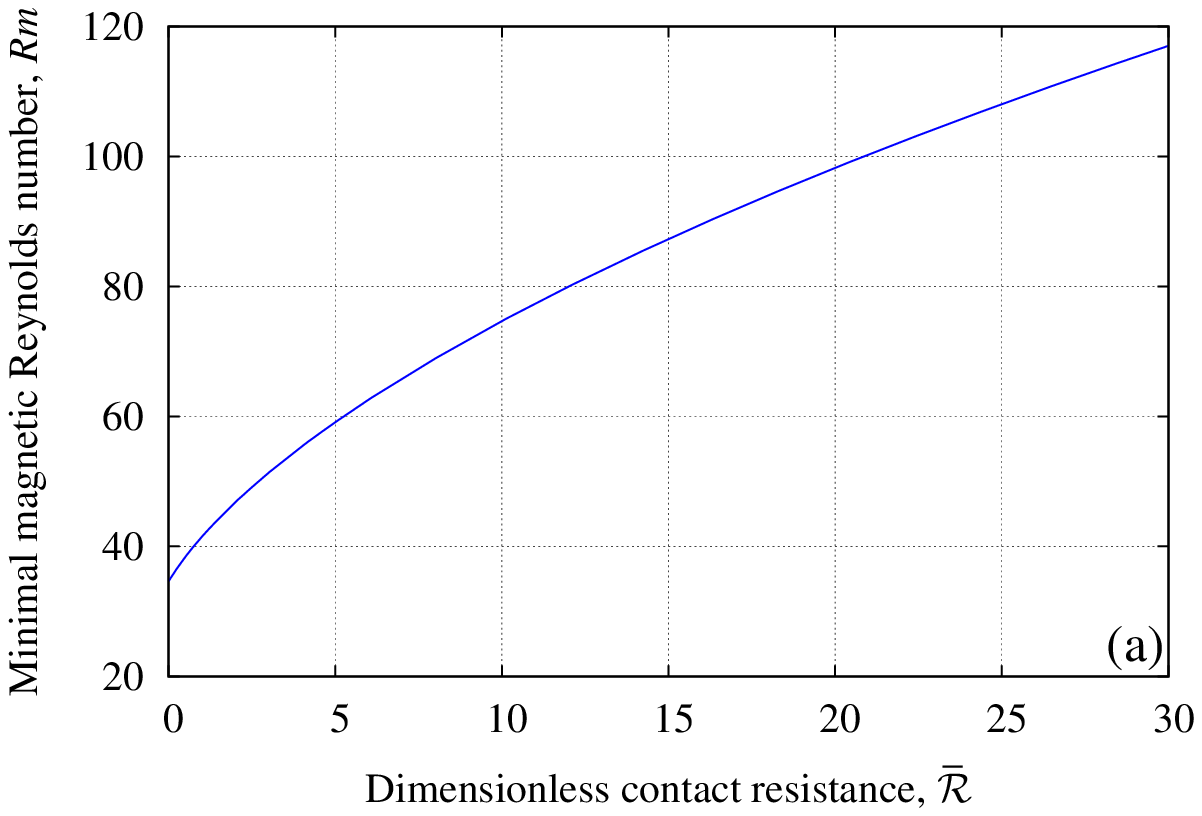}\includegraphics[width=0.5\textwidth]{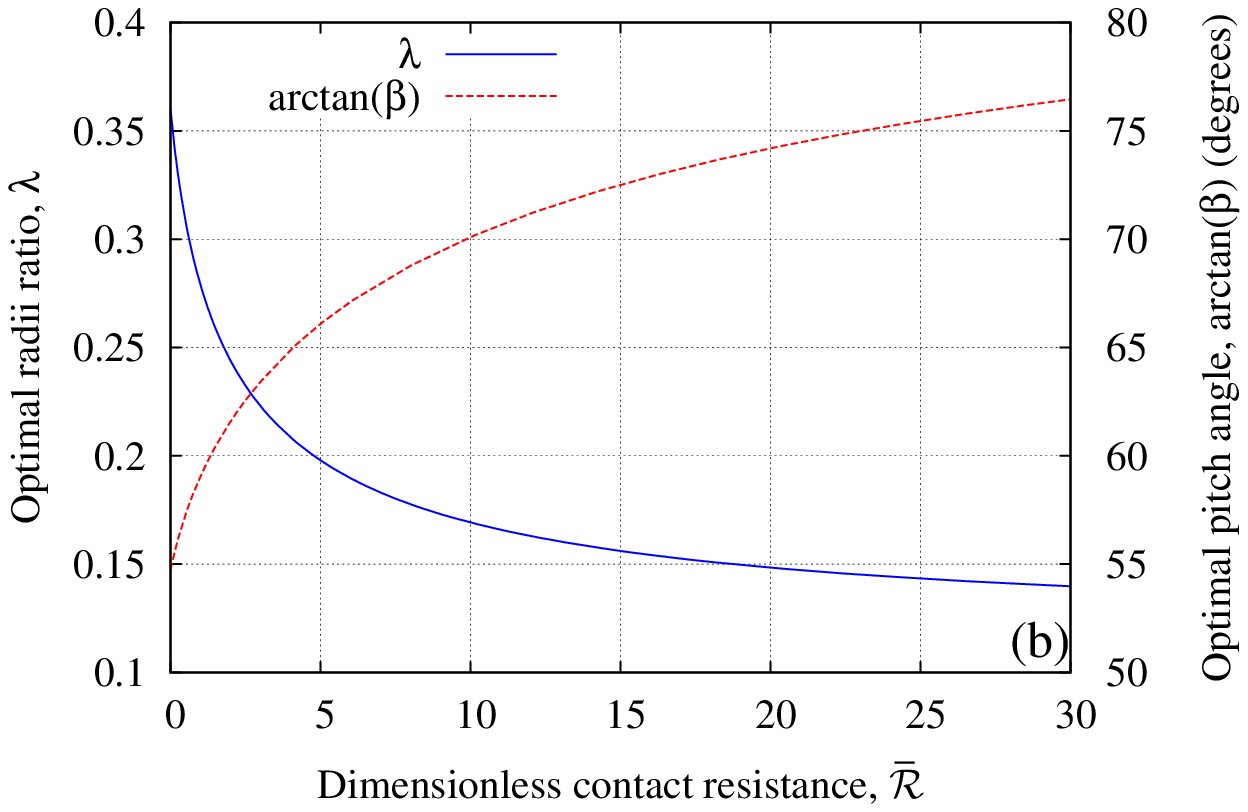} 
\par\end{centering}

\caption{\label{fig:crtpar}Minimal magnetic Reynolds number $\Rm$ (a), optimal
radii ratio $\lambda,$ and the coil pitch angle $\arctan\beta$ (b)
versus the dimensionless contact resistance$\bar{\mathcal{R}}.$ }
\end{figure*}
The case of solid rotating ring considered above corresponds to $\kappa=1,$
whereas $\kappa=2$ corresponds to the rotating ring sectioned similarly
to the stationary one except for the opposite direction of the spiral
slits. As seen from the expression above, the latter case is equivalent
to the former with both $\Rm$ and $\beta$ reduced by a factor of
$\sqrt{2}.$

Now, let us determine the optimal $\beta$ and $\lambda$ that yield
the lowest $\Rm$ for a given $\bar{\mathcal{R}}.$ In the simplest
case of a negligible contact resistance, which corresponds to $\bar{\mathcal{R}}=0,$
Eq. (\ref{eq:Rm}) yields $\Rm\sim2/(\kappa\beta)+\beta.$ It means
that $\Rm$ attains a minimum at $\beta_{c}=\sqrt{2/\kappa},$ which
corresponds to the optimal pitch angles of $54.74^{\circ}$ and $45^{\circ}$
for only one and both rings sectioned. The respective lowest values
of $\Rm$,  $34.63$ and $24.49$, are attained at the same optimal
radii ratio of $\lambda=0.3602.$ (see Fig. \ref{fig:Rm-lmb} for
the case of only one ring sectioned). The minimal $\Rm$ increases
with  $\bar{\mathcal{R}},$ which also causes a steep reduction of
the optimal radii ratio and a comparably fast rise of the pitch angle
(see Fig. \ref{fig:crtpar}).

\section{Feasible setup}

Finally, let us evaluate the rotation rate required for self-excitation
in a setup with the outer radius of $R_{o}=\unit[30]{cm}$ and the
ring thickness of $d=\unit[3]{cm.}$ First, we need to estimate electrical
resistance of sliding liquid-metal contacts. A suitable metal for
such contacts may be the eutectic alloy of GaInSn \citep{Maribo-etal10},
which is liquid at room temperature with the kinematic viscosity $\nu=\unit[3.5\times10^{-7}]{m/s^{2}},$
electrical conductivity $\sigma_{\mathrm{GaInSn}}=\unit[3.3\times10^{6}]{S/m}$
\citep{Mue-Bue}. Assuming the contact gap width of $\delta=\unit[0.5]{cm}$
and the inner radius $R_{i}\approx\unit[10]{cm},$ we have $\mathcal{R}_{i}\approx\frac{\delta\sigma_{\mathrm{GaInSn}}^{-1}}{2\pi dR_{i}}\approx\unit[0.02]{\mu\Omega.}$
The resistance of the outer contact is by a factor of $\lambda=R_{i}/R_{0}=0.33$
lower than $\mathcal{R}_{i}.$ Then the dimensionless contact resistance
can be estimated as 
\[
\bar{\mathcal{R}}=2\pi\sigma_{\mathrm{Cu}}d\mathcal{R}_{i}(1+\lambda)\approx\frac{\sigma_{\mathrm{Cu}}}{\sigma_{\mathrm{GaInSn}}}\frac{\delta}{R_{i}}(1+\lambda)\approx0.2
\]
If only one disc is sectioned, which is easier to manufacture, the
respective magnetic Reynolds number in Fig. \ref{fig:crtpar}(a) is
$\Rm\approx40.$ This corresponds to the rotation frequency $f=\frac{\Omega}{2\pi}=\frac{\Rm}{2\pi\mu_{0}\sigma_{\mathrm{Cu}}dR_{o}}\approx\unit[10]{Hz,}$
which is well within the operation range of standard AC electric motors.
The respective linear velocity of the outer edge of the ring is around
$v\approx\unit[20]{m/s}.$ At this velocity the tensile stress at
the rim of the ring, $\rho_{\mathrm{Cu}}v^{2}\approx\unit[4]{MPa},$
is more than by an order of magnitude below the yield strength of
annealed Copper \citep{Li-Zinkle12}. The optimal inner radius $R_{i}\approx0.3R_{0}\approx\unit[9]{cm}$
following from Fig. \ref{fig:crtpar}(b) is not far from the value
assumed above. The respective pitch angle for $\beta\approx1.6$ is
about $58^{\circ}.$ 

The number of spiral arms is determined by the following arguments.
The current distribution defined by Eq. (\ref{eq:J-phi}) can hold
only in the inner parts of the ring which are radially confined between
the spiral slits. This ideal distribution is expected to break down
at the rims, which are radially exposed to the edges of the ring located
at the nearly equipotential metal liquid contacts. In order to confine
this perturbation to the outer rim with $r/R_{o}\gtrsim0.9,$ Eq.
(\ref{eq:spiral}) suggests that $\frac{-2\pi}{\beta\ln0.9}\approx40$
equally distributed spiral slits are required.

The last critical issue is the viscous power losses associated with
the turbulent drag acting on the outer sliding contacts at high shear
rates. These losses can be estimated as $Q=S\tau v\approx\unit[7]{kW},$
where $S\approx2\pi dR_{o}$ the area of the outer sliding contact,
$\tau=\frac{c}{4}\frac{\rho v^{2}}{2}$ is the turbulent shear stress,
and $c\approx0.02$ is the Darcy friction factor for turbulent pipe
flow with the Reynolds number $\textit{Re}\sim10^{5}$ \citep{MarSon87}. 

In conclusion, the proposed disc dynamo design appears feasible in
terms of both the disc spinning rate and the power required to drive
it. Note that the relatively large setup size is due to the turbulent
energy dissipation which scales as $Q\sim(dR_{o})^{-2}\sim\Omega^{2}.$
Namely, reducing the system size by one third would require about
five times higher power input to achieve self-exitation of the magnetic
field.

\section*{Acknowledgment}

R.A.-Z. is grateful to the National Council for Science and Technology
of Mexico (CONACYT) for funding the projects CB-168850 \& 131399.

\balance

\end{document}